\documentclass[twocolumn,prl,amsmath,amssymb,showpacs,superscriptaddress]{revtex4-1}
\usepackage{epsf}      
\usepackage{graphicx}
\usepackage{color}
\usepackage{gensymb}
\usepackage[dvipsnames]{xcolor}
\usepackage{amsmath}
\usepackage[outercaption]{sidecap} 

\begin{document}

\title{Frustrated magnetism in antiferromagnetic nonsymmorphic square-net lattice: NdSbSe}

\address{Department of Chemistry, Indian Institute of Technology Delhi, New Delhi 110016, India}
\author{Prabuddha Kant Mishra}
\address{Department of Chemistry, Indian Institute of Technology Delhi, New Delhi 110016, India}
\author{Priyanka Nehla}
\address{Department of Materials and Engineering, \AA ngstr\"{o}m Laboratory, Uppsala University, Box 35 75103, Sweden.}
\address{Centre for Neutron Scattering, \AA ngstr\"{o}m Laboratory, Uppsala University, Box 35 75103, Sweden.}
\address{National Forensic Sciences University, Goa, 403401, India.}
\author{Rishabh Shukla}
\address{Laboratory for High-Performance Ceramics, Swiss Federal Laboratories for Materials Science and Technology (Empa), Überlandstrasse 129, 8600 Dübendorf, Switzerland}
\author{Rie Umetsu}
\address{Institute for Materials Research, Tohoku University, 2-1-1 Katahira, Aoba-ku, Sendai 980-8577, Japan}
\author{Ashok Kumar Ganguli}\email[E-mail: ]{ashok@chemistry.iitd.ac.in}
\address{Department of Chemistry, Indian Institute of Technology Delhi, New Delhi 110016, India}
\address{Department of Chemical Sciences, Indian Institute of Science Education and Research, Berhampur, Odisha-760003, India}

\begin{abstract}
Spintronics has emerged as a field of vast applicability, and layered magnetic materials have served as a ground for advancement in this direction. Here, we report the synthesis and detailed magnetic and specific heat studies on NdSbSe, (a ZrSiS-based structure) magnetic topological material. The temperature-dependent magnetization shows the presence of competing magnetic interactions ($T_N<T<150$ K) in addition to a long-range antiferromagnetic (AFM) ordering below 4.2~K. In the AFM state, the isothermal magnetization confirms spin reorientation to the critical magnetic field of 40 kOe. Frequency-dependent ac-susceptibility measurements have probed the nonequilibrium dynamics of frustrated magnetic moments (near 150 K). The $\lambda$-like peak at 3.8~K observed in the specific heat shifts to a lower temperature with applied magnetic fields and validates the AFM order. In addition, the specific heat does not exhibit any sign corresponding to the short-range magnetic order near 150 K (spin-glass-like memory effect). In addition, the derived parameters from specific heat suggest the presence of a strong electronic correlation in NdSbSe, resulting in a Kondo-like signature in temperature-dependent resistivity data. %Moreover, the existence of crystal field splitting of $f$-electronic states has been evidenced by analysis of inverse susceptibility vs. temperature plots and a hump in the specific heat between 5--20 K.

\end{abstract}

%\date{\today}

%\pacs{74.70.Xa}

\maketitle

%%%%%%%%%%%%%%%%%%%%%%%%%%%%%%%
\section{Introduction}

The role of magnetic spin in technological advancement has been guiding the field of spintronics. 
Colossal magnetoresistance in correlation with non-symmorphic crystal symmetry has been evident from experimental investigations on ZrSiS and other nonmagnetic members of this family, specifying them as nodal-line semimetal \cite {Neupane2016, Topp2016, Schoop2016, Fu2019, Ali2016, Singha2017, Song2022}.
In the recent past, the ZrSiS-type layered magnetic material has been studied to understand and tune the interplay between topology and magnetism \cite{Salters2023, Kirby2023, Lei2021, Dalgaard2020, Ke2020, Chikina2023}. Due to the nonsymmorphic crystallographic symmetry, the square-net-like geometry of ZrSiS structures is known to generate a nontrivial band structure with symmetry-protected four-fold Dirac states in the Brillouin zone \cite{Klemenz2020, Klemenz2019}. 
In addition, the theoretical studies on nonmagnetic LaSbTe (ZrSiS type), predicted weak topological states \cite{Qiunan2015} with considerable spin-orbit coupling and verified by experimental results suggesting Dirac-like dispersion as a nontrivial characteristic signature of the electronic band structure in LaSbTe \cite{Singha2017a}.

Confirmation of non-trivial topological electronic band structure has been provided by angle-resolved photoemission spectroscopy (ARPES) for GdSbTe \cite{Hosen2018} and CeSbTe \cite{Schoop2018}. Similarly, the first-principle studies and the magnetotransport experiments on HoSbTe infer the existence of nontriviality \cite{Yang2020}. In addition to nontrivial topology, isostructural magnetic analogues have been investigated for other characteristics like charge density wave, discrete magnetization (devil's staircase) \cite{Sankar2020, Pandey2020, Chen2017, Regmi2022, Gao2022, Yang2020}. Further, the inevitable crystal field splitting of energy levels due to the preferred direction of stacking leads to additional features like magnetic frustration as observed in members with late rare-earth magnetic elements as competitive short-range magnetism along with long-range ordering of moments in HoSbTe and TbSbTe \cite{Plokhikh2022}. The long-range antiferromagnetic coupling of moments is commonly observed for all compositions containing $f$-electrons. In rare-earth-based ZrSiS-type systems, most of the attention on physical characterization is limited to telluride members mainly, while interesting aspects related to the role of isovalent substitution on fermi energy tuning and magnetic structure need to be explored. To the best of our knowledge, the only selenium analogues of LnSbTe that have been examined for their magnetic properties are LaSbSe \cite{Pandey2022}, CeSbSe  \cite{Chen2017, Singha2021} and GdSbSe \cite{Gautam2024}. However, a comprehensive magnetotransport study for CeSbSe is yet to be explored.
By comparing the $\gamma$ for these compounds, we found that the electronic correlation is enhanced from CeSbTe ($\gamma$ = 41 mJ K$^{-2}$ mol$^{-1}$) to CeSbSe ($\gamma$ = 8.6 mJ K$^{-2}$ mol$^{-1}$) \cite{lv2019, Chen2017}. Due to the correlated physics of electron filling, Fermi surface modulation and competing magnetism, the role of substitution is crucial to explore the interplay among them.

In particular, magnetic studies on NdSbTe have shown interesting results of change in the easy axis of AFM ordering with gradual substitution of Sb with Te \cite{Karki2024, Salters2023}. The study has motivated the investigation of the nature of magnetic interactions in the Se-analogue of NdSbTe. Although the crystal structure of NdSbSe has been known for a long the physical characteristics of this layered ZrSiS-type material are yet to be explored.

In this study, we have successfully synthesized polycrystalline NdSbSe through solid-state reaction. Detailed magnetic measurements have been carried out to explore ground-state interactions of NdSbSe. Interestingly, we have observed a signature of magnetic frustration in the system along with robust long-rang antiferro-type magnetic ordering. To the best of our knowledge, NdSbSe is the first early rare-earth-based compound to exhibit frustrated magnetic interactions. Eventually, the increase in the applied field suppressed this particular signature of magnetic frustration. The isothermal magnetic measurements indicate a reorientation of moments with field at low temperatures. Furthermore, parameters obtained from the field-dependent specific heat studies yield details of thermal parameters and establish NdSbSe as a strongly correlated electronic system.

%%%%%%%%%%%%%%%%%%%%%%%%%%%%%%%%
\begin{figure*}
\includegraphics[width= 2.0\columnwidth,angle=0,clip=true]{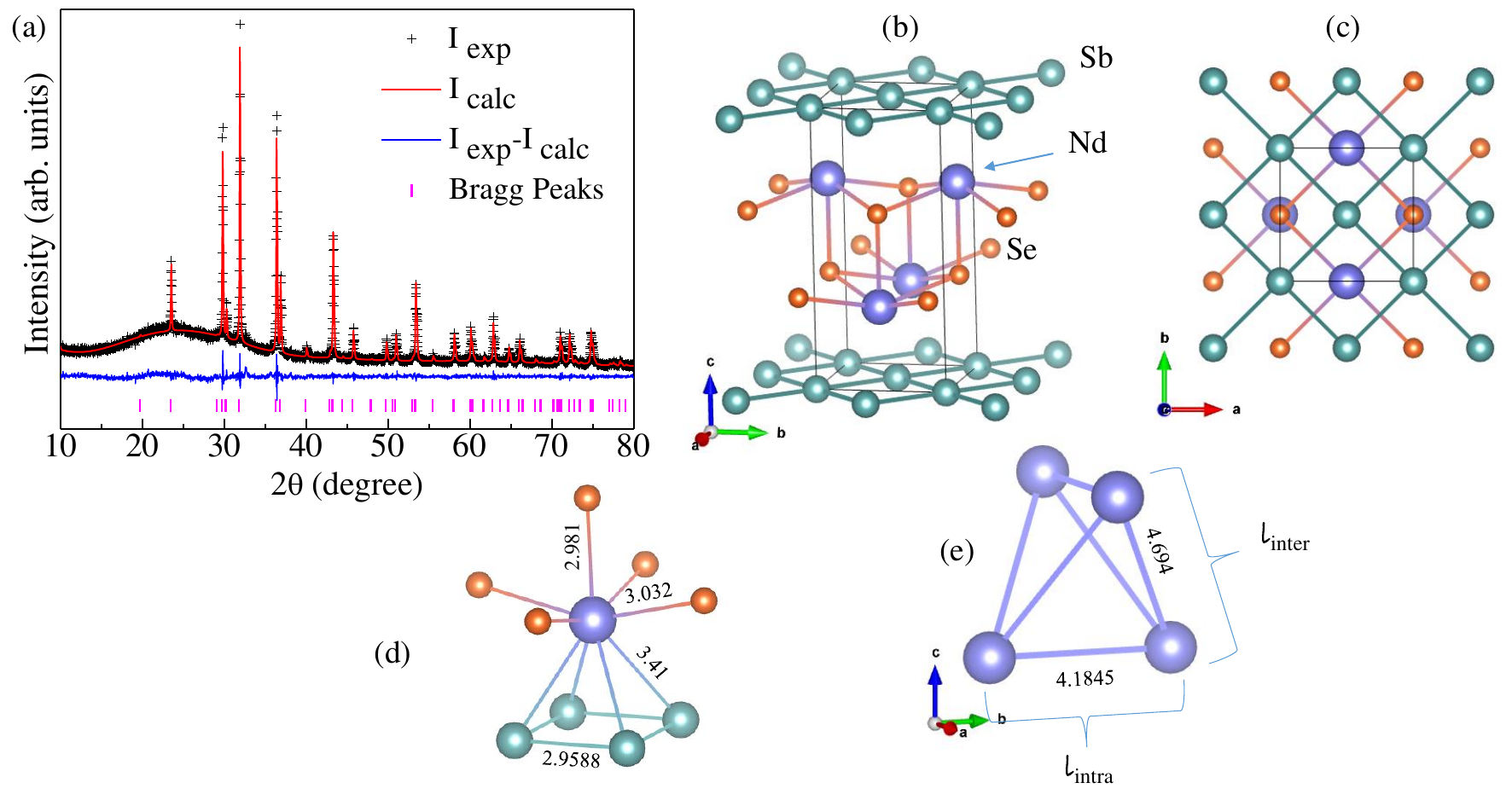}
\caption{(Color online) (a) Rietveld refinement of room temperature powder X-ray diffraction pattern of polycrystalline NdSbSe. The black cross and solid red lines show the experimental and calculated intensity pattern. Vertical bars indicate the allowed Bragg's reflections for NdSbSe. The blue line indicates the difference between the experimental and calculated patterns. (b) shows the schematic crystal structure of NdSbSe. (c) The schematic crystal view along the c-axis depicts the square-net arrangement, (d) The polyhedral unit around Nd ion. (e) The distorted tetrahedral arrangement of Nd in the lattice, with two distinct types of bonds.}
\label{Fig1}
\end{figure*}

%%%%%%%%%%%%%%%%%%%%%%%%%%%%%%%

\section{Experimental Details}

Polycrystalline NdSbSe has been synthesized by the sealed tube method. The elemental forms of Nd, Sb, and Se have been ground together in the stoichiometric ratio (Nd:Sb:Se::1:1:1) inside an argon-filled glove box and sealed in an evacuated quartz tube for heat treatment. A reaction condition of 1150~K for 48 hours was used inside a box furnace. The product (black-coloured powder) obtained from the first heating was ground well, pelletized, and sintered at 1150 K for 48 hours to achieve better phase homogeneity and pellet-shape stability. The phase of the final sintered material (in the powder form) has been characterized using X-ray diffraction technique using Bruker D8 Advance diffractometer with Cu-K$_{\alpha}$ radiation performed on powder as shown in \ref{Fig1}(a). The structural refinement on powder X-ray diffraction data was carried out using the Rietveld method with the TOPAS software package \cite{Topas}. Energy dispersive X-ray analysis (EDX) to accomplish the homogeneity and compositional analysis. The temperature and field-dependent magnetic measurements have been performed in a Superconducting Quantum Interference Device (SQUID), from Quantum Design, USA. The specific heat was measured as a function of temperature and applied field in a Physical Property Measurement System (PPMS), from Quantum Design, USA.

%%%%%%%%%%%%%%%%%%%%%%%%%%%%%%%%%%%%%%%%%%%%%%%%%%
\begin{table}[h]
\scriptsize\addtolength{\tabcolsep}{-1pt}
%\scriptsize%\addtolength{\tabcolsep}{-5pt}
%\caption{\label{table1}Structural parameters}
\caption{\label{table1}Structural and magnetic parameters of NdSbSe in their respective units.}

\begin{ruledtabular}
\begin{tabular}{p{3.2cm}p{1.5cm}p{0.3cm}p{0.5cm}p{1.2cm}p{1cm}}
NdSbSe \\
Space group: & $P4/nmm$ \\
Space group number: & 129\\
{\it a}(\r{A}): & 4.1845(1) \\
{\it c}(\r{A}): & 9.0190(2) \\
\hline

Atom & Site & x & y & z & Occu.\\
Nd & 2c & 0 & 1/2 & 0.2980(3) & 1.01(1)\\
Sb & 2a & 0 & 0 & 0 & 0.98(1) \\
Se & 2c & 0 & 1/2 & 0.6286(2) & 1.02(1)\\
%\end{tabular}
\hline

$T_N$ (K) & 4.2 $\pm$ 0.1 \\
$\theta_P$ (K) & -25 $\pm$ 1 \\
$C$ (emu Oe$^{-1}$mol$^{-1}$K$^{-1}$ ) & 1.70$\pm$0.2 \\
$\mu_{eff}^{theo}$ ($\mu_B$/Nd$^{+3}$) & 3.62  \\
$\mu_{eff}^{calc}$ ($\mu_B$/Nd$^{+3}$) & 3.75$\pm$0.05 \\
$M_S$ at 2 K (emu g$^{-1}$) & 9.2$\pm$0.1 \\
$H_C$ at 2 K (Oe) & 10$\pm$5 \\

\end{tabular}
\end{ruledtabular}
\end{table}

%%%%%%%%%%%%%%%%%%%%%%%%%%%%%%%%%%%%%%%%%%%%%%%%%%%%%%%%

\section{Results and Discussion}

\subsection{Structural Analysis}

Figure~\ref{Fig1}(a) shows the Rietveld refinement of room temperature powder X-ray diffraction pattern of polycrystalline NdSbSe. It crystallizes into a ZrSiS-type nonsymmorphic tetragonal crystal structure with space group $P4/nmm$ ($\#129$) and obtained lattice parameters are $a$ = 4.1845(1)~\AA~and $c$ = 9.0190(2)~\AA \cite{Pandey2022}. The details of the refined structural parameters are summarized in Table~\ref{table1}. Further, Fig.~\ref{Fig1}(b) depicts the schematic crystal structure of NdSbSe (created using VESTA software \cite{Momma2008}), where Nd and Se occupy the 2c sites while 2a sites are filled with Sb, and each Nd-ion is coordinated by the five Se and four Sb-ions. The structure can be viewed as an alternate stacking of Sb and Nd-Se layers along the $c$-axis of the lattice. The square-net arrangement is depicted in Fig.~\ref{Fig1}(c). In polyhedra around Nd, two distinct Nd-Se bonds can be observed (Fig. \ref{Fig1}(d)). The four in-plane ($\parallel$ ab plane) Nd-Se bonds are equivalent and slightly longer (3.032 \r{A}) than one out-of-plane Nd-Se bond (2.981 \r{A}) i.e. $\parallel$ to c-axis.

Considering the earlier explored NdSbTe, we have compared the lattice for substituting chalcogenides. Modulation in bond lengths through substitution has a significant role in magnetic interactions. The Nd-Nd interatomic distances ($l_{Nd-Nd}$) are crucial to consider for understanding magnetic properties. In the unit cell, two distinct Nd-Nd interatomic distances are realized (1) intralayer ($l_{intra}$) and (2) interlayer one ($l_{inter}$). Moving from composition NdSbTe to NdSbSe, the $l_{intra}$ changes from 4.35 to 4.18 \r{A} and the $l_{inter}$ changes from 5.37 to 4.69 \r{A}. It suggests a significant suppression in $l_{inter}$ compared to $l_{intra}$. In the case of NdSbSe, the minimal difference in interatomic distances in the distorted tetrahedral geometry (Fig.~\ref{Fig1}(e) fuels competition among interlayer and intralayer magnetic interaction and causes frustrated magnetism \cite{Volkova2006, Milam2019}.

Another approach to quantifying the 2D nature of electronic systems is associated with the tolerance factor in square net materials \cite{Klemenz2020}. The tolerance factor is defined as $t= l_{sq}/ l_{nn}$, where $l_{sq}$ and $l_{nn}$ are interatomic distances in square-net and nearest neighbour distances from the square-net. For NdSbSe, $t$ is found to be 0.9, which is similar to NdSbTe (0.9) \cite{Sankar2017} and GdSbSe (0.87) \cite{Gautam2024}. Thus NdSbSe belongs to the same structural class of ZrSiS-type layered materials having ionic interaction among adjacent layers \cite{Gebauer2021}. In literature, tolerance factor has been used for the classification of materials through cutoff value, and materials having $t \leq $ 0.95 are expected to have an inverted band structure while materials having $t \geq $ 1.05 are anticipated to have a non-inverted band structure \cite{Klemenz2020}.
Thus, NdSbSe is expected to have nontrivial topology in electronic states, like realized in NdSbTe \cite{Regmi2023}. Further, the delocalized bonding in the square-net of Sb is key to stabilizing the square-net geometry and contributes to transport properties \cite{Klemenz2019, Klemenz2020, Schoop2018}.

%%%%%%%%%%%%%%%%%%%%%%%%%%%%%%%%%%%%%%%%%%%%%%%%%%%%%
%%%%%%%%%%%%%%%%%%%%%%%%%%%%%%%%%%%%%%%
\begin{figure*}[t!]
\includegraphics[width=2.0 \columnwidth,angle=0,clip=true]{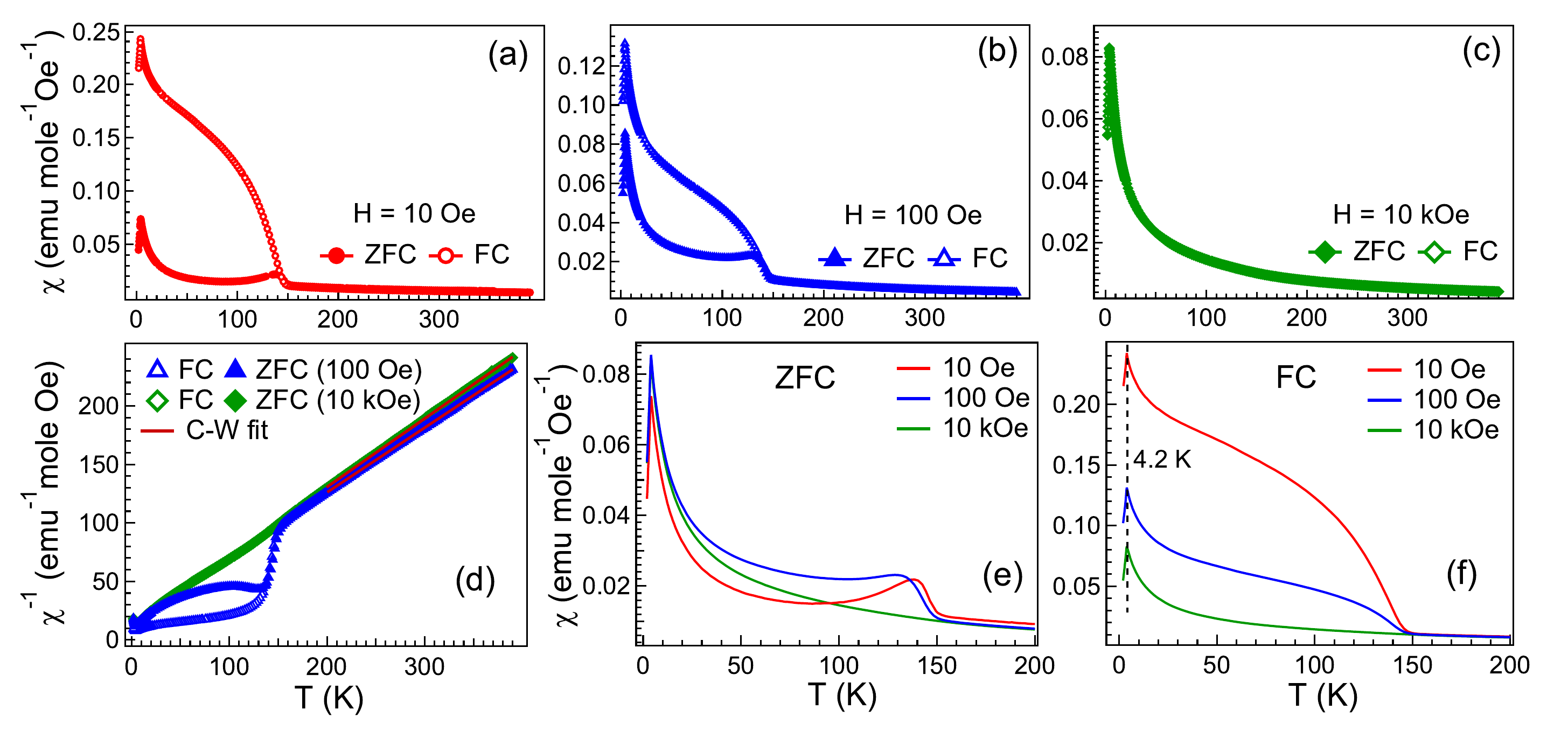}
\caption{(Color online) The dc-magnetic susceptibility as a function of temperature in the temperature range of 2-400 K under zero-field-cooling (ZFC) and field-cooling (FC) protocol at an applied field of (a) 10 Oe, (b) 100 Oe, and (c) 10 kOe. (d) Temperature dependence of $\chi^{-1}(T)$ and the red solid line shows linear CW fit in the temperature range of 200-390 K. The dc-magnetic susceptibility as a function of temperature in (e) zero-field-cooling and (f) field cooling protocols for various applied fields of 10 Oe, 100 Oe and 10 kOe.}
\label{Fig2}
\end{figure*}

%%%%%%%%%%%%%%%%%%%%%%%%%%%%%%%%%%%%%%%%%%%%%
\subsection{Magnetization Measurements}

The dc-magnetization has been measured as a function of temperature from 2--390 K at various applied magnetic fields between 10~Oe to 10~kOe, using ZFC (zero-field cooling) and FC (field cooling) measurement protocols. Figures~\ref{Fig2}{(a--c)} show the temperature-dependence of dc-magnetic susceptibility ($\chi_{dc}$) at applied magnetic fields of H = 10~Oe (red), 100~Oe (blue), and 10~kOe (green), respectively. Here two distinct magnetic transitions are evident from $\chi_{dc}$ vs. T graphs. For a clear description, we have described the magnetic behavior in two temperature regimes separately as (i) high-temperature regime (150~K $<T<$ 390~K) and (ii) low-temperature regime (2~K $<T<$ 150~K). In the (i) high-temperature regime, the ZFC and FC magnetization overlap, and $\chi_{dc}$ increases with a decrease in the temperature similar to the trivial paramagnetic behavior. Therefore, the inversely proportional temperature dependence of $\chi_{dc}$(T) (Fig.~\ref{Fig2}{(d)}) was analyzed with the Curie-Weiss (C-W) law in the paramagnetic region given as \cite{Chen2017},

\begin{equation}\label{CW}
\chi_{dc} (T) = \frac{N_A\mu_{eff}^2}{3k_B(T-\theta_p)}
\end{equation}
\\
where $N_A$, $k_B$, and $\mu_{eff}$ are Avogadro's number, Boltzmann constant, and effective magnetic moment, respectively. In the C-W equation, $\theta_p$ is the Weiss temperature signifying the ground state of magnetic interaction (+ve/-ve for ferromagnetic/nonferromagnetic interactions). In Fig.~\ref{Fig2}{(d)}, the $\chi^{-1}_{dc}$(T) vs. $T$ plots are best fitted (solid red line) using the C-W model in a high-temperature regime of 390~K to 200~K. From the best fits in Fig.~\ref{Fig2}{(d)}, $\mu_{eff}$ is estimated to be 3.75(5)~$\mu_B/Nd^{+3}$ ions, and $\theta_p$ is -25(1)~K, which suggests antiferromagnetic type exchange interactions among moments. The calculated value of $\mu_{eff}$ is consistent with the theoretical calculated value ($g\sqrt{J(J+1)}\mu_B$) of 3.62~$\mu_B/Nd^{+3}$. Interestingly, in the (ii) low-temperature regime, the $\chi^{-1}_{dc}(T)$ data shows a considerable deviation from linearity (C-W fit) ($T<$ 150 K), suggesting existing magnetic interactions in a close resemblance with the magnetic properties of CeSbSe \cite{Chen2017} and CeSbTe \cite{Schoop2018}. The observed deviation below identical temperature ($T<$ 150 K) for NdSbSe and CeSb(Se/Te) suggests the existence of interactions with similar energy scales in these structures, possibly the crystal field splitting of $f$-electronic levels \cite{Chen2017}. In Figs.~\ref{Fig2}{(a-c)}, we observe a bifurcation between ZFC and FC curves near 150~K (T$_{irr}$), and a peak in the ZFC curve near 150~K is highlighted in Fig.~\ref{Fig2}(e), which is usually observed in glassy-spin systems \cite{Shukla2023}. In addition, a flattening and shift of this peak towards a lower temperature with an increase in the applied magnetic field further confirm the glassy-spin behavior in NdSbSe (discussed below in detail). This characteristic behavior is absent in $\chi_{dc}$ (T) measured at an applied field of 10~kOe as shown in Fig.~\ref{Fig2}(e) with a solid green line, indicating suppression of the short-range magnetic ordering with higher applied magnetic fields. In addition, a long-range magnetic ordering of moments with antiferromagnetic (AFM) interactions occurs at Neel temperature ($T_N$) of 4.2~K, as evident from the Fig.\ref{Fig2}(a-c) and is also highlighted in Fig.~\ref{Fig2}(f). This AFM transition is robust for applied magnetic fields and does not shift with the applied magnetic fields of 10~kOe (see dotted black line in Fig.~\ref{Fig2}(f)). It is worth mentioning, that the AFM-like long-range ordering has been explored for isostructural materials like NdSbTe and CeSbTe\cite{Chen2017, Sankar2020, Pandey2020, Gautam2024, Mishra2024NdBiTe}. Next, we have calculated the frustration index ($f=\mid\theta_P\mid/T_N$) $\approx$6 for NdSbSe from magnetization measurement. The frustration index is almost twice larger than that observed in the case of NdSbTe \cite{Sankar2020}. It suggests the presence of considerable magnetic frustrated interactions in NdSbSe and aligns with observed bifurcation in magnetization.

\subsection{Magnetic Memory Measurements}
%%%%%%%%%%%%%%%%%%%%%%%%%%%%%%%%%%%%%%%%%
\begin{figure}[t!]
\includegraphics[width= 1.0\columnwidth,angle=0,clip=true]{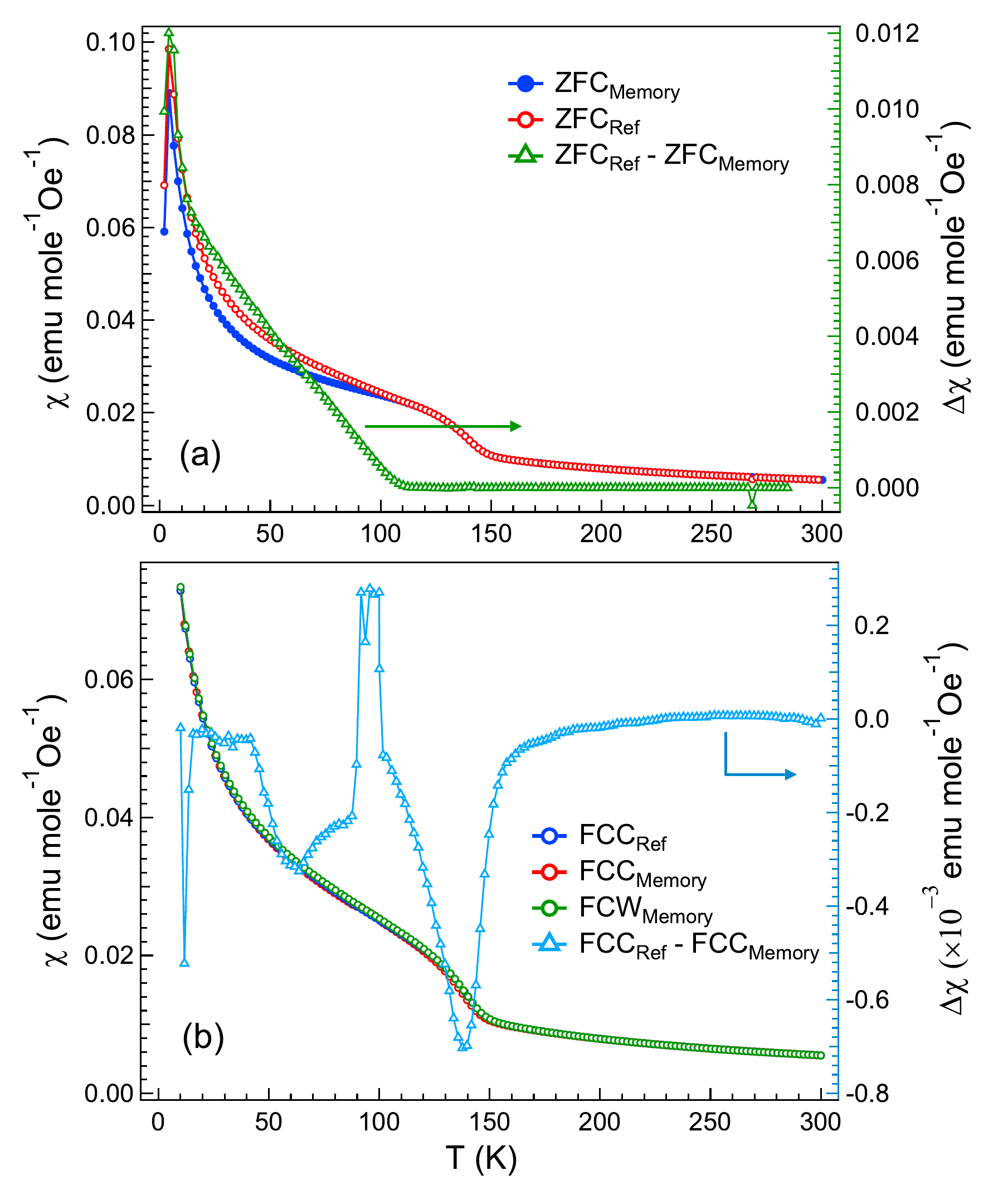}
\caption{(Color online) The temperature dependence of dc-magnetic susceptibility measured under (a) ZFC and (b) FC protocols designed to probe memory effect, with a waiting time of 1 hour at 100 K. The difference between memory and reference curves is shown on the right scale.}
\label{Fig3}
\end{figure}
%%%%%%%%%%%%%%%%%%%%%%%%%%%%%%%%%%%%%%%%%

To further confirm and analyze the glass-like short-range interactions, the dc-magnetic memory effect measurements are performed between 2-300~K under the ZFC and FC memory protocols. The slow spin dynamics having out-of-equilibrium responses in magnetization measurements can be explored as a memory effect that is greatly affected by the magnetic and thermal history of the material. In this experiment, the sample is cooled from the paramagnetic region to a temperature below the $T_{irr}$ in the desired cooling protocol of ZFC or FC, and a dwell for a certain time is applied below $T_{irr}$ followed by further cooling to 2~K. The dc-magnetic data were recorded during the cooling/warming of the sample with an applied magnetic field. A data set without a halt is used as a reference curve and a baseline to identify a clear change in the magnetization at the waiting temperature under the memory effect. The memory ZFC and FC measurements were performed with a dwell time of 1 hour at 100~K ($<T_{irr}$). In Fig.~\ref{Fig3}(a), the open red circles and solid blue circles show the ZFC curve recorded during the warming of the sample without (reference curve) and with a dwell time of 1 hour at 100~K (memory curve) . The difference of the memory and reference curve (open green triangles) on the right-axis of Fig.~\ref{Fig3}(a) shows a clear deviation near 100~K corresponding to the dwell temperature during the memory measurement, which confirms the presence of a short-range magnetic ordering in NdSbSe below $T_{irr}$. Similarly, the memory effect measurement is performed under the FC protocol and shown in Fig.~\ref{Fig3}(b). In Fig.~\ref{Fig3}(b), the open blue, red, and green circles manifest the field cooling reference curve, field cooling memory curve under cooling, and FC-magnetization measured during warming. Since under the FC protocol, the magnetic field is applied during both cooling and warming and the magnetic field is not switched off during dwell time of 1 hour at 100~K ($<T_{irr}$), therefore to highlight the memory effect, the difference of FCC reference and memory effect (open indigo triangles) is shown on the right axis of Fig.~\ref{Fig3}(b), where an abrupt change in the magnetization can be seen at dwell temperature, indicating a memory effect due to the presence of frustrated magnetic interactions under a low applied magnetic field. Therefore, from Figs.~\ref{Fig3}(a,b) the memory effect further validates the presence of short-range magnetic ordering in the NdSbSe. The magnetic frustration in ZrSiS type of materials has been found in the heavier rare-earth analogues \cite{Plokhikh2022}. However, the possible signatures of memory effect are yet unexplored in this class (ZrSiS) of materials.

%%%%%%%%%%%%%%%%%%%%%%%%%%%%%%%%%%%%%%%%%%%%
\begin{figure*}[t!]
\centering
\includegraphics[width= 2.0\columnwidth,angle=0,clip=true]{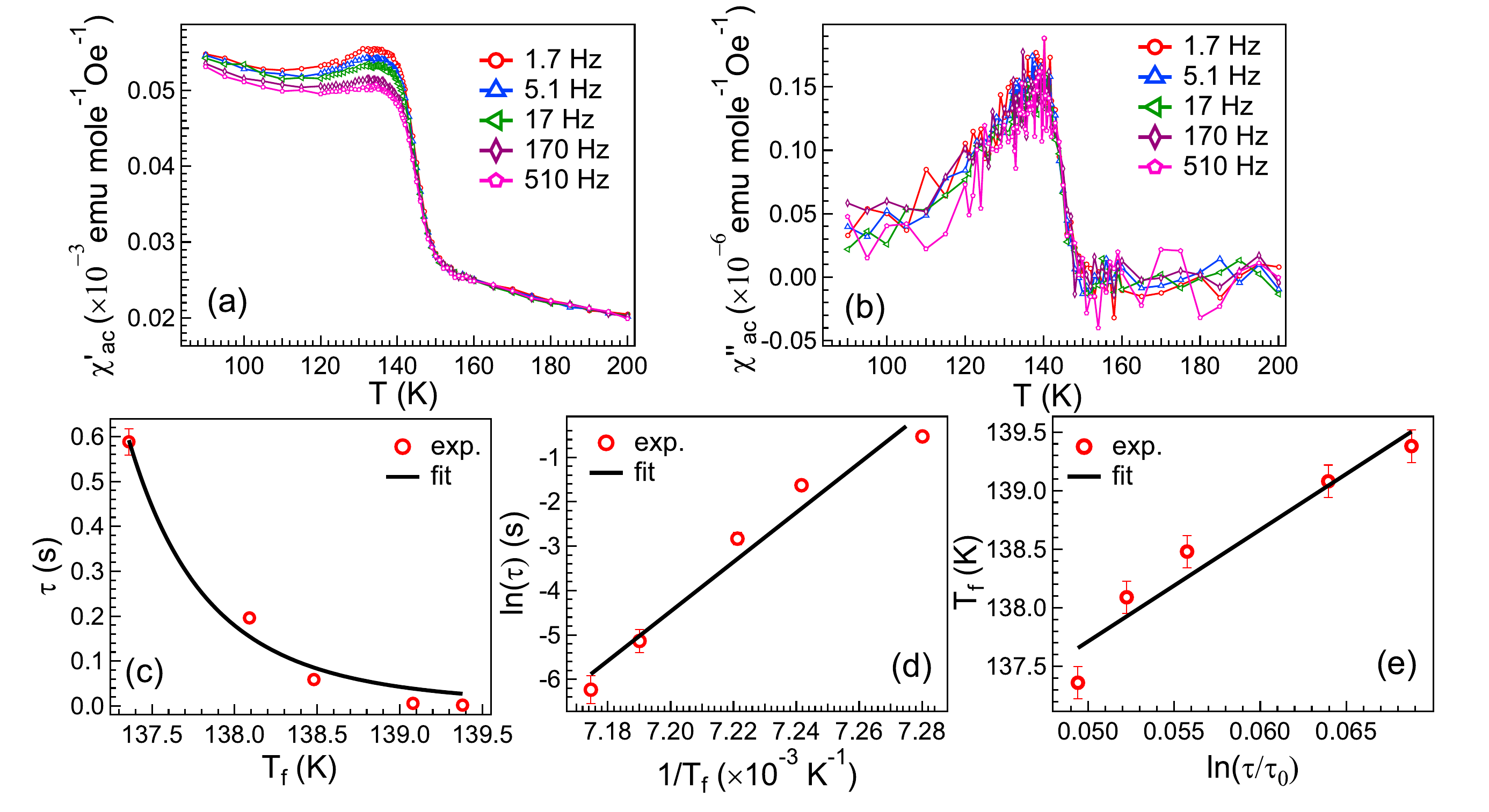}
\caption{(Color online) (a) In-phase ($\chi'_{ac})$ and (b) out-of-phase  ($\chi''_{ac})$ data of ac-susceptibility as a function of temperature at various excitation frequencies. The best fit curves for (c) power law and (d, e) Vogel-Fulcher law.}
\label{Fig4}
\end{figure*}

%%%%%%%%%%%%%%%%%%%%%%%%%%%%%%%%%%%%%%

\subsection{ac-magnetic susceptibility studies}

The irreversibility in the ZFC and FC magnetization data has suggested the presence of competing short-range magnetic interactions in NdSbSe and validated under the memory effect measurements (discussed above). To further explore the origin of short-range magnetic interactions close to $T_{irr}$ (150~K), temperature-dependent ac-magnetic susceptibility ($\chi_{ac}$) measurements have been performed to probe the spin dynamics with temperature and excitation frequency. The temperature-dependent in-phase ($\chi'_{ac}$) and out-of-phase ($\chi''_{ac}$) components of ac-magnetic susceptibility at different excitation frequencies of 1.7~Hz to 510~Hz are shown in Figs.~\ref{Fig4}(a,b), respectively. The peak-shaped anomaly in Figs.~\ref{Fig4}(a,b) is observed close to $T_{irr}$ (150~K) in both $\chi'_{ac}$ and $\chi''_{ac}$ shows the presence of glassy magnetic state as expected earlier. The temperature dependence of peak position on the applied excitation frequency has been further analyzed to understand the spin dynamics in NdSbSe. The peak observed in $\chi_{ac}$ shifts to a lower temperature and reduces in amplitude with increased excitation frequency similar to spin-glass compounds. This shift in the peak position (T$_f$) with excitation frequency is first analyzed by the Mydosh parameter \cite{Mydosh1993} ($S$), defined as

\begin{eqnarray}\label{mydosh}
S = \frac{\Delta T_f}{T_f \Delta log_{10}(f)}
\end{eqnarray}

where T$_f$, and $f$ are the freezing temperature and excitation frequency, respectively. The $S$ parameter is calculated from the endpoints of the lowest and highest measured excitation frequencies of 1.7~Hz and 510~Hz and results in a value of 0.006. The obtained value of $S$ is slightly higher as compared to the values reported for the canonical spin glasses, 0.0045 (for AuMn) and 0.004 (for CuMn) \cite{Mydosh1993} and close to the values reported for the cluster spin glasses (0.008-0.018) \cite{Chakrabarty2014, Shukla2019}. Since $S$ is an endpoint measure, therefore, to estimate the time scale of spin dynamic relaxation and magnetic interaction between spin moments in short-range magnetic order, the frequency dependence of freezing temperature (T$_f$) is analyzed with power law \cite{Souletie1985}, expressed as
\begin{eqnarray}
\tau = \tau_0 \left(\frac{T_f-T_{SG}}{T_{SG}}\right)^{-z\nu'}
\end{eqnarray}
where $\tau$ ($f=$ 1/$\tau$), $\tau_0$, T$\rm_{SG}$ are the relaxation time, characteristic relaxation time for flipping of single spin entity, and spin-glass temperature corresponding to divergence of $\tau$, respectively. Also, z$\nu'$ (lies between 4 and 12) and $\nu'$ are the dynamic critical exponent and critical exponent associated with the spin correlation length $\xi$ = (T$_f$/T$_{SG}$-1)$^{\nu'}$. Figure~\ref{Fig4}(c) shows the best fit (solid black line) to the experimental data using power law and yields the values of $\tau_0$ = 3.5$\pm$0.3$\times$10$^{-10}$~s, T$\rm_{SG}$ = 134.8$\pm$0.5~K, and z$\nu'$ = 5.4$\pm$0.3. The $\tau_0$ exists in the range of 10$^{-12}$ to 10$^{-14}$~s and 10$^{-6}$~s to 10$^{-10}$~s for canonical and cluster spin glasses, respectively \cite{Shukla2019, Anand2012}. For NdSbSe the obtained value of $\tau_0$ confirms the presence of cluster spin glass and indicates the slower spin dynamics due to interactions between the spin entities. 

Further confirmation of the cluster spin glass in NdSbSe is investigated from the frequency dependence of $\tau$ by the Arrhenius relation, $\tau = \tau_0 exp(\frac{-E_a}{k_BT})$. The failure of the Arrhenius relation (not shown here) manifests the presence of magnetic interaction between the individual spins, i.e., cluster spin glass. In order to further analyze the spin dynamics, Arrhenius law is further modified to the Vogel-Fulcher law given below \cite{Mydosh1993, Shukla2019, Souletie1985},

\begin{eqnarray}
\tau = \tau_0 exp\left(\frac{-E_a}{k_B(T_f-T_0)}\right)
\end{eqnarray}

where T$_0$ is the Vogel-Fulcher temperature gives an estimate of the strength of magnetic interaction between spins and other parameters are the same as defined earlier. The Vogel-Fulcher law is further modified to, $ln(\tau) = ln(\tau_0)-\frac{E_a/k_B}{T_f-T_0}$, which is used to plot a graph between ln($\tau$) vs. 1/T$_f$ in Fig.~\ref{Fig4}(d). A linear fit (solid black line) to the experimental data in Fig.~\ref{Fig4}(d) yields the values of T$_0$ = 134.7$\pm$0.3 and $\tau_0$ = 2.8$\pm$0.5$\times$10$^{-9}$~s. Also, another simplified form of Vogel-Fulcher law, $T_f =\frac{E_a/k_B}{ln(\tau/\tau_0)}+T_f$ is used to plot a graph between T$_f$ vs. ln($\tau/\tau_0$) in Fig.~\ref{Fig4}(e), and the best fit to the experimental data (solid black line) yields the values of T$_0$= 132.9$\pm$0.9 and E$_a$= 8.2$\pm$0.6~meV. Here, a non-zero value of T$_0$ (comparable to the freezing temperature, T$\rm_f$) in the present sample further validates the interaction between the magnetic clusters \cite{Anand2012}.
%%%%%%%%%%%%%%%%%%%%%%%%%%%%%%%%%%
\begin{figure*}[t!]
\centering
\includegraphics[width= 2.0\columnwidth,angle=0,clip=true]{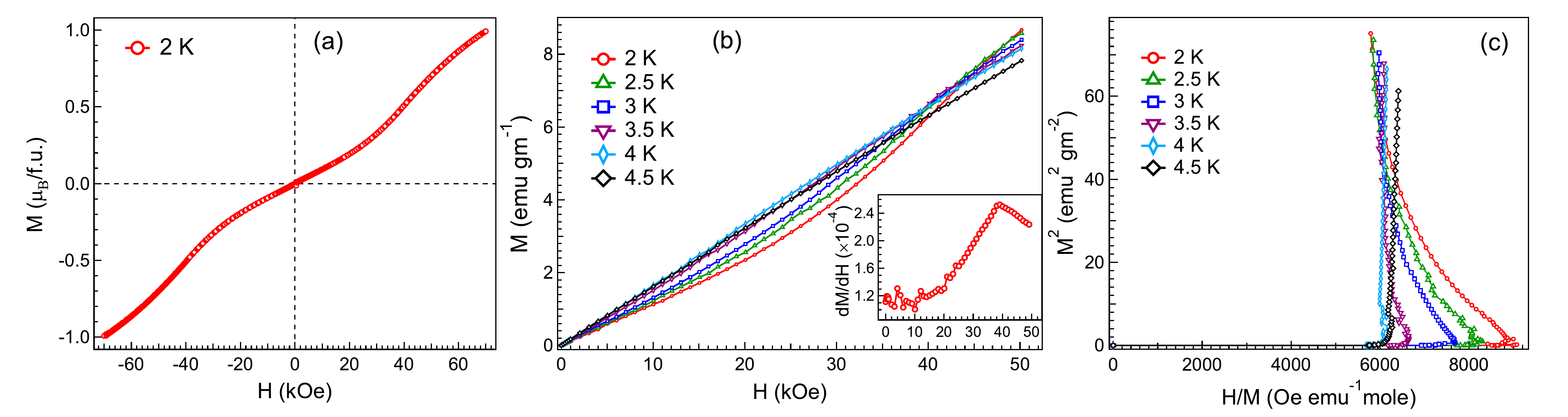}
\caption{(Color online) (a) Isothermal magnetization $M(H)$ measurement at 2 K in the temperature range of $\pm$ 70 kOe. (b) The $M-H$ curves at various temperatures across the transition temperature. The inset shows the derivative plot $dM/dH$ vs $H$ for $T$= 2 K, having a peak at the field for spin reorientation. (c) The Arrott plot ($M^2$ $vs.$ $H/M$) in the range of 2 K to 4.5 K with $\Delta T$ = 0.5 K.}
\label{Fig5}
\end{figure*}
%%%%%%%%%%%%%%%%%%%%%%%%%%%%%%%%%%

\subsection{Isothermal Magnetization}

Isothermal magnetic measurements of NdSbSe have been performed at T= 2~K below $T_N$ within the applied magnetic field of $\pm$70 kOe, as shown in Fig.~\ref{Fig5}(a). The absence of hysteresis in the $M(H)$ loop is consistent with the antiferromagnetic state. Interestingly, magnetization changes the slope with an increase in the magnetic field, suggesting a spin-reorientation type metamagnetic transition and commonly observed in other isostructural antiferromagnetic materials \cite{Pandey2020, Schoop2018, lv2019, Lei2019, Gao2022}. The derivative of magnetization with respect to the applied field ($dM/dH$) for $T=$ 2 K shows a peak at $H_{Cr}$ = 42 kOe, as shown in the inset of Fig.~\ref{Fig5}(b). The transition from AFM state to canted-AFM state, where the canting of spins leads to deviation from linearity in the isotherms and adds a ferromagnetic-like response in the $M(H)$ data. In the canted state ($H>$42 kOe), the observed magnetization for applied field 70 kOe at 2 K is 1 $\mu_B/Nd^{+3}$, considerably lower than the theoretical value of saturated magnetic moment (3.28 $\mu_B/ Nd^{+3}$) for free ion and comparable with the reported value for NdSbTe \cite{Pandey2020}. Such reduced magnetization can be related to crystal field splitting of energy states and considerable single ion anisotropy of Nd$^{+3}$ \cite{Gao2022, Sankar2020}. The virgin plots recorded at various temperatures around the $T_N$ have been used to find the Arrott plot, $M^2$ $vs.$ $H/M$ and shown in Fig.~\ref{Fig5}(c). The negative slope of curves in Arrott plots suggests a first-order transition from paramagnetic to AFM state, supported by Banerjee criterion \cite{Banerjee1964}.

%%%%%%%%%%%%%%%%%%%%%%%%%%%%%%%

%%%%%%%%%%%%%%%%%%%%%%%%%%%%%%%%%%%%%%%%%

\subsection{Specific Heat Measurements}

\begin{figure}[t!]
\includegraphics[width= 1.0\columnwidth,angle=0,clip=true]{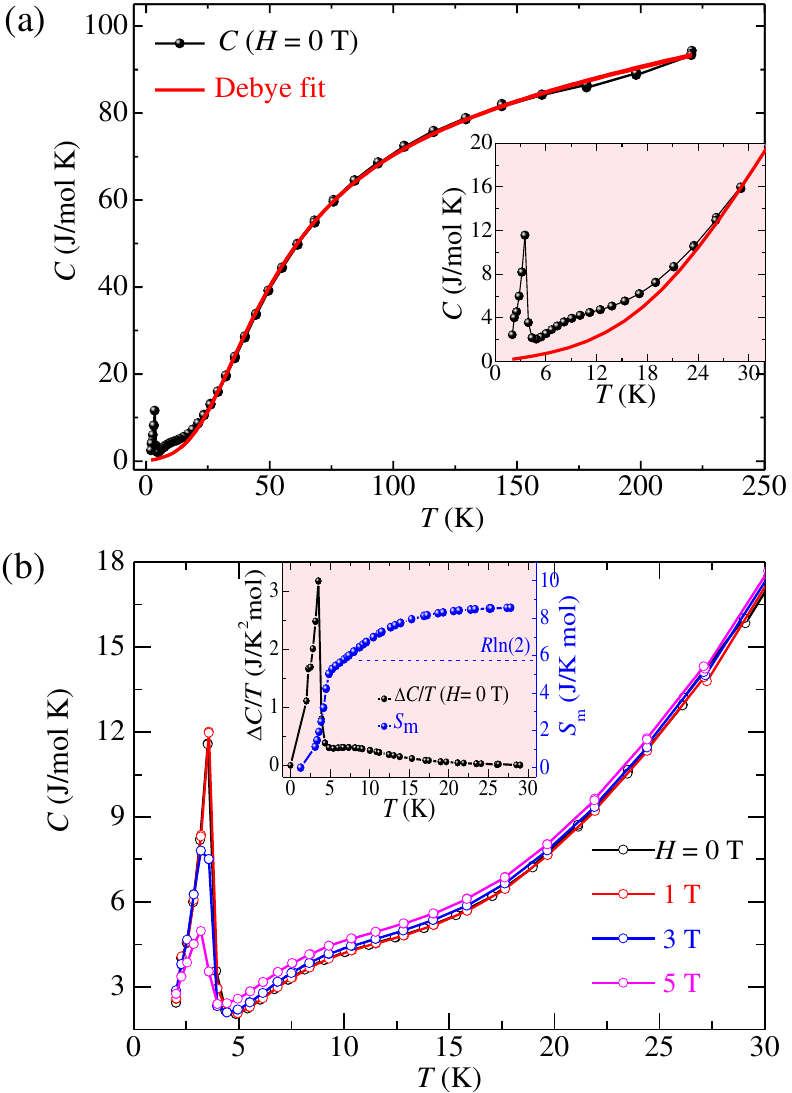}
\caption{(Color online) (a) Specific heat ($C$) data as a function of temperature in the range of 2-220 K in the absence of magnetic field, fit with the Debye equation is shown by the solid red line. The inset shows data with the enlarged view in the temperature range of 2-30 K. (b) The $C(T)$ data under various applied fields for $H \leq 50$ kOe, inset shows $\Delta C/T$ vs. $T$ and magnetic entropy in the temperature range 2-30 K. More details are mentioned in the text.}
\label{Fig6}
\end{figure}

The specific heat ($C$) measurements have been performed under various applied magnetic fields to gain a detailed understanding of existing magnetic interactions. The specific heat data have been shown in Fig.~\ref{Fig6}(a) in the temperature range of 2-220 K at $H$= 0 T. The $\lambda$-peak has been observed at 3.8 K, confirming the long-range antiferromagnetic ordering consistent with the temperature-dependent $\chi_{dc}$ data. The measured $C(T)$ data contains contributions from electronic, lattice and magnetic interactions. These various contributions can be extracted from the data by having an idea about their temperature dependence and the temperature region in which they contribute. In the high-temperature range ($T>>T_N$), $C(T)$ data can be analyzed with the Debye equation, without considering magnetic contribution \cite{Chen2017, Gao2022}. The combined equation $C$= $\gamma T$ + $C_{Debye}$ can be written as

\begin{equation}\label{CT}
C(T) =  \gamma T + \alpha 9 n R \left( \frac{T}{\Theta_D} \right)^3 \int^{(\Theta_D/T)}_T \frac {x^4e^x}{(e^x-1)^2} dx 
\end{equation}

where $\gamma$ and $\Theta_D$ are the Sommerfeld coefficient and Debye temperature respectively. The deviation from fit, in the lower temperature regime (shown in the inset of Fig.~\ref{Fig6}(a)) originates from the magnetic contribution. From the fit $\gamma$ and $\Theta_D$ are estimated to be 258(2) mJ K$^{-2}$ mol$^{-1}$ and 164.5$\pm$ 8 K, respectively. The relatively large value of $\gamma$ as compared to the isostructural antiferromagnets (115 mJ/K$^2$ mol for NdSbTe) \cite{Pandey2020} and much larger as compared to nonmagnetic materials (2.19 and 0.51 mJ/K$^2$ mol for LaSbSe and LaSbTe respectively) \cite{Pandey2022}, suggests enhanced electronic correlations in the NdSbSe, having contribution from Nd-4$f$ electrons. The possibility of Kondo hybridization between conducting and localized $f$ electronic states results in mass enhancement, which is evident in other isostructural magnetic systems \cite{Pandey2020, Regmi2022, lv2019}. The Sommerfeld coefficient, $\gamma$ is related to the density of states $N(E_F)$ of the system close to the Fermi level, which is calculated by using the relation $\gamma = [\pi^2 k_B^2 N(E_F)]/3.$ Thus $N(E_F)$ is found to be $\approx$ 14 states/eV f.u. Next, the obtained $\theta_D$ for NdSbSe is lower than that of NdSbTe ($\theta_D$= 236 K), indicating weaker interatomic interactions present in NdSbSe as compared to NdSbTe. A similar trend with the substitution has been reported for nonmagnetic analogues LaSb(Se/Te) \cite{Pandey2022}. While an opposite trend in the value of $\theta_D$ on substitution of Te with Se is reported on isostructural ZrSi(S/Se/Te) materials \cite{Sankar2017, Song2021}. The phonon-specific heat coefficient, $\beta$, is related to the Debye temperature of the material by the relation $\Theta_D = (\frac{12}{5\beta}\pi^4nN_Ak_B)^{1/3}$, where $n$ is the number of atoms in the formula unit, $N_A$ is the Avogadro number, and $k_B$ is the Boltzmann constant. $\beta$ is estimated to be 0.348(5) mJ K$^{-4}$ mol$^{-1}$. A field-dependent specific heat data is shown in the inset of Fig.~\ref{Fig6}(a), it is evident that the intensity of the anomaly decreases and shifts to a lower temperature with the increase in the applied magnetic field, a usual signature of AFM interaction. A significant hump in $C_P$ data in the temperature range of 5-15 K is likely to be associated with the crystal field splitting, which is present for all applied fields. 
%%%%%%%%%%%%%%%%%%%%%%%%%%%%%%%%%%%%%%%%%

\begin{figure}[t!]
\includegraphics[width= 1.0\columnwidth,angle=0,clip=true]{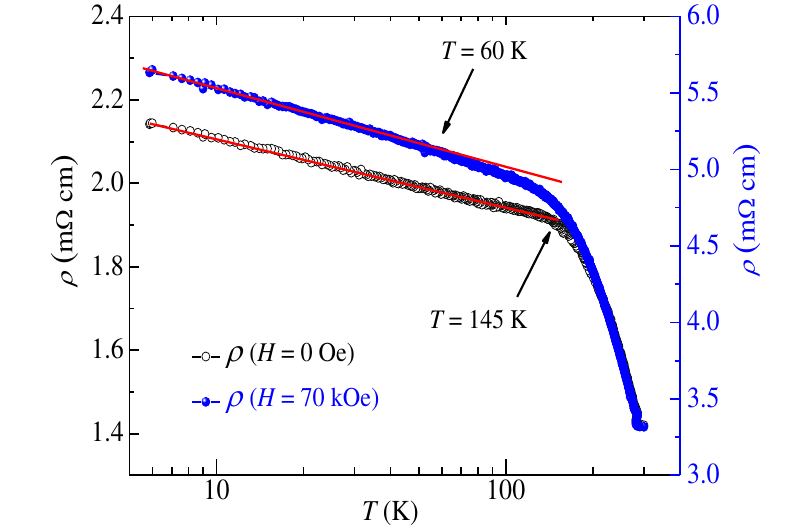}
\caption{(Color online) Resistivity as a function of temperature in the range of 5-300 K on a semi-logarithmic temperature scale, for the applied field of 0 and 70 kOe. The red solid line shows a linear best fit and suggests Kondo-interaction.}
\label{Fig7}
\end{figure}

%%%%%%%%%%%%%%%%%%%%%%%%%%%%%%%%%%%%
The magnetic contribution to specific heat ($C_{mag}$) has been extracted by the subtraction of the fitting curve from the total experimental $C$ data, as presented in the inset of \ref{Fig7}(b). Further, the estimated magnetic entropy can be approximated by using $S_m = \int_0^T \frac{C_m(T)}{T}dT$ and was found to be 8.5 J/K mol up to 30 K, which is almost 73$\%$ of the theoretical value of $S_m$ ($R$ln$(2S+1)$ = 11.53 J/K mol), where $R$ is the universal gas constant and $S = 3/2$ for Nd$^{+3}$ ion. The recovered reduced $S_m$ is comparable to that observed for NdSbTe \cite{Sankar2020} and suggests the existence of the crystal electric field effect (CEF). Moreover, the calculated S$_m$ approaching $R$ln$(2S+1)$ suggests that the CEF of Nd$^{+3}$ is a $\Gamma$8 quartet as the ground state \cite{Sankar2020}.
%%%%%%%%%%%%%%%%%%%%%%%%%%%%%%%%%%%%%%%%%%%

%%%%%%%%%%%%%%%%%%%%%%%%%%%%%%%%%%%%%%%%%%%%%%%%%%%
%%%%%%%%%%%%%%%%%%%%%%%%%%%%%%%%%%%%%%%%%%%%%
\subsection{Temperature-dependent Resistivity}

The temperature dependence of resistivity exhibits nonmetallic behaviour and rises upon cooling as shown in Fig.~\ref{Fig7}. More specifically the resistivity varies linearly with the logarithmic scale of temperature and archives value of 2.15 m$\Omega$cm at 5 K. Similar to $\chi_{dc}(T)$ data, the resistivity shows two distinct temperature dependence and can be separated in low (5 K$<T<$150 K) and high (150 K$<T<$300 K) temperature regimes broadly. As shown in Fig.~\ref{Fig7}, the resistivity rises steeply in the temperature range of 300-150 K, while it increases with a reduced rate at lower temperatures $T<$ 150 K. As indicated in Fig.~\ref{Fig7}, the change in the slope of resistivity is concomitant with the peak observed in the ac-susceptibility data. Under the applied field (70 kOe), the deviation from logarithmic dependence is observed at lower temperatures ($T=$ 60 K). Such a logarithmic temperature dependence is an indication of Kondo-like behaviour and consistent with observed large $\gamma$ from specific heat data, which also has been observed in NdSbTe \cite{Pandey2020}. The LnSbTe compounds have metal-like electronic band structures that still have shown nonmetallic transport \cite{lv2019, Pandey2020}. Though the variation of resistivity looks similar to NdSbTe, the interplay among the magnetism and topology is different, resulting in positive magnetoresistance. The positive magnetoresistance observed in NdSbSe distinct from the negative MR reported in NdSbTe \cite{Pandey2020} suggests a change in the transport mechanism realized through the compositional variation. A complete comprehensive magnetotransport and Hall measurements on a single crystal for probing the topology is desirable and under progress. 
%%%%%%%%%%%%%%%%%%%%%%%%%%%%%%%%%%%%%%%%%%%

\section{CONCLUSIONS}

We have synthesized polycrystalline NdSbSe and observed long-range antiferromagnetic (AFM) transition, with $T_N$ near 4 K from temperature-dependent magnetic and specific heat measurements. The $T_N$ is largely unaffected by the increase in the field strength, suggesting the robust nature of interactions. The observed bifurcation in ZFC and FC magnetization data below 150~K ($T_{irr}$), suggests the existence of short-range magnetic interactions, which have been suppressed gradually for higher applied fields. Consistent with the bifurcation, the memory effect in magnetization has been observed under both ZFC and FC protocols. Finally, the presence of short-range interaction (cluster spin glass ) has been confirmed by ac-susceptibility measurements performed at various excitation frequencies. The structural insight indicates the existence of competing exchange interactions due to the comparable interlayer and intralayer interatomic distances and crystal field splitting of $f$-electronic states. Further, the magnetic entropy associated with AFM order is calculated from specific heat data and suggests the existence of crystal electric field splitting. In the AFM state, the magnetic field-driven metamagnetic transition is evident from isothermal magnetization data. The Kondo interaction among the conducting electrons and localized moments is supported by the observation of large $\gamma$ in specific heat and logarithmic temperature dependence of resistivity. From these experimental investigations, NdSbSe has been established as a layered nonsymmorphic antiferromagnet with strongly correlated electrons. Due to the presence of crystal field effect, expected nontrivial topology, and long-range magnetic order in the layered lattice, NdSbSe is an interesting material to be explored for spintronic applications.

%%%%%%%%%%%%%%%%%%%%%%%%%%%%%%%%%%%%%%%%%%%%

%\section*{ACKNOWLEDGMENTS}

%The authors thank CRF, IIT Delhi for SQUID facility. We thank Prof. S. Patnaik for priliminary transport measurement. MN acknowledges MHRD, India for fellowship through IIT Delhi. SA acknowledges DST, India for fellowship through INST Mohali. CKV acknowledges CSIR, India for senior research fellowship (grant no.: 09/086(1297)/2017-EMR-I). AKG thanks SERB-DST, Government of India for financial support (sanction no.: EMR/2016/000156). Computational results are based on the computations using the High Performance Computing cluster, Padum, at IIT Delhi.
\section{\noindent ~Acknowledgments}
AKG thanks SERB, Government of India for funding. PKM acknowledges the Council of Scientific $\&$ Industrial Research (CSIR), [09/086(1425)/2019-EMR-I] India for fellowship. PN acknowledges the financial support from Swedish Agency for Economic and Regional Growth under the project SSRE. The transport measurements have been performed at IMR-Tohoku, under GIMRT program (202312-HMKPA-0504).

\bibliographystyle{apsrev4-2}
\bibliography{NdSbSe}

\end{document}